\documentclass[12pt]{article}
\title{ INSTANTON SCREENING IN THE NONPERTURBATIVE GLUODYNAMICS }
 \author{ N.O.Agasian \\
 Institute of Theoretical and Experimental Physics \\ 117218,
 Moscow, B.Cheremushkinskaya 25, Russia\\
e-mail:agasyan@vxitep.itep.ru}
\date{}
\newcommand{\be}{\begin{equation}}
 \newcommand{\ee}{\end{equation}}
\begin{document}
\maketitle
\begin{abstract}

\noindent
The gluon fields screening in the stochastic vacuum of gluodynamics is
studied. The effective action is derived for the instanton interacting with
nonperturbative fields. Quantum nonperturbative effects are shown to affect
greatly  the shape of instanton. The power asymptotics $x^{-2}$ of
the classical "instanton's profile function" at large
distances is replaced due to these effects by Airy function asymptotics.
 \end{abstract}

\section{Introduction}

In the last years a systematic description of nonperturbative effects in QCD
has been provided in terms of the gluon nonlocal gauge--invariant
correlators [1].
This  Vacuum Correlators Method turned out to be very successful
in phenomenological description of important QCD phenomena  [2].
By now, there exist numerical results from lattice simulations concerning the
fundamental field strength correlators for pure gauge theory with gauge group
SU(2) [3] and SU(3) [4] over physical distances ranging up to O(1) fm.

On the other hand during the last 18 years we have witnessed active
development of nonperturbative QCD picture based on the  instanton liquid
model [5,6,7].  It is therefore of great interest to compare the lattice
calculations of field strength correlators with the computation of the same
quantities in instanton--anti--instanton vacuum model.
The analytical  calculations of field strength correlators for the
instanton--anti--instanton vacuum, at least, in the dilute gas approximation 
exhibit the power decrease of the bilocal correlator at large distances. On 
the other hand, the results of lattice simulations are consistent with more 
rapidly decrease of this correlator. Thus we see that the computation of the 
field strength  correlator requires the comprehensive analysis of instanton 
field character in nonperturbative vacuum. It is necessary to note that 
already in first publications on the instanton liquid model the Debye 
screening was revealed with a typical mass of the order of 350 MeV [6].  A 
new mechanism of instanton scale stabilization essentially different from 
that considered in [6] was proposed in [8,9]. It is due to the effective 
"freezing" of strong coupling constant in the nonperturbative vacuum.

In the present paper are  obtained the effective action of topologically
nontrivial fluctuations in the nonperturbative vacuum making use of the
Vacuum Correlators Method. The new type is revealed of instanton screening
different from previously considered [6,10].

\section{Effective action}

The Euclidean action of gluodynamics is written as
\be
S=\frac{1}{2g^2}\int d^4x tr {\cal F}^2_{\mu\nu}(x),
\ee
where we use the Hermitian matrix form for gauge fields
\be
{\cal
F}_{\mu\nu}=\partial_{\mu}{\cal
A}_{\nu}-\partial_{\nu}{\cal A}_{\mu}-i[{\cal A}_{\mu},{\cal
A}_{\nu}], ~~{\cal A}_{\mu}=g{\cal A}^a_{\mu}t^a,~~
trt^at^b=\frac{1}{2}\delta^{ab}
 \ee
  To obtain effective action for
instanton in the nonperturbative vacuum we represent the full field
${\cal A}_{\mu} $
 as
  \be
  {\cal  A}_{\mu}=A_\mu+B_\mu,
  \ee
   where $A_\mu$  is
the external field (instanton--like object with unit topological
charge) and $B_{\mu}$  are nonperturbative vacuum fields.

Under gauge transformations ($U$ is the transformation matrix), the fields
change as
$$
A_\mu(x)\to U^+(x)A_{\mu}  (x)U(x)
$$
\be
B_\mu\to U^+(x)(B_\mu(x)-i\partial_{\mu})U(x)
\ee

The general expression for the effective action of an instanton in the
nonperturbative vacuum has the form
\be
Z=e^{-S_{eff}[A]} = <e^{-S[A+B]+S[B]}>_B,
\ee
where
\be
<\hat 0(B)>_B=\int d\mu [B]\hat 0(B)
\ee
and $d\mu{[B]}$ is  the measure of  integration with respect to
nonperturbative fields. The
explicit form of this measure will not be needed in our analysis.
The full field strength can be rewritten in the form
\be
{\cal  F}_{\mu\nu}[A+B]=F_{\mu\nu}[A]+G_{\mu\nu}[B]-i[A_{\mu},
B_{\nu}]-i[B_{\mu},A_{\nu}]
\ee
Let us now consider a large--size instanton. The squared strength of the
instanton field at the center is given by
\be
(F_{\mu\nu}^{inst}(x=x_0))^2=192/\rho^4
\ee
For $\rho>\rho_0    =(192/<G^2>)^{1/4}\sim 1 fm$, the instanton field is weak
relative to the characteristic field  strengths in  the vacuum gluon
condensate \\
 $<G^2>\equiv <(gG^a_{\mu\nu})^2> \simeq 0.5 Gev^4$ [11] and can
be treated as a perturbation.
Therefore, expanding the effective action  (5)
up  to the second--order terms in
$A_{\mu}$ we have
\be
S_{eff}[A]=-lnZ\simeq S[A]+<S_{dia}[A,B]>_B+<S_{para}[A,B]>_B,
\ee
where
\be
S_{dia}{[A,B]}=-\frac{1}{2g^2}\int
d^4xtr\{[A_{\mu},B_{\nu}]+[B_{\mu},A_{\nu}]\}^2,
\ee
\be
S_{para}{[A,B]}=-\frac{1}{4g^2}\int
d^4xd^4y tr F_{\mu\nu}(x) G_{\mu\nu}(x) tr
F_{\sigma\lambda}(y) G_{\sigma\lambda} (y),
\ee
and the conditions $<B_{\mu}>=0, <G_{\mu\nu}>=0$ are taken into account.

Hence, there exist  two contributions [9].
First, there is the motion of "charged particles" in the external
gluonic field $F_{\mu\nu}$. This motion leads to screening similar to
orbital Landau diamagnetism. And second, there is the direct  interaction of
$F_{\mu\nu}$ with the spin of the field $B_{\mu}$. In the second order in the
external field, this interaction yields an antiscreening term, which is
analogous to that describing the Pauli paramagnetic effect.

A similar decomposition into diamagnetic and paramagnetic terms was suggested
by Polyakov [12] for perturbative gluodynamics, and led to the simple and
elegant explanation of antiscreening with  correct coefficient for
$\beta$--function.

 Let us consider paramagnetic component of $S_{eff}$.  The
general form of the instanton field is
 \be
 A_{\mu}(x)=\Phi(x,x_0)\Omega
\bar A_{\mu}(x-x_0)) \Omega^+ \Phi (x_0,x),
 \ee
 where
 \be \Phi(x,x_0)=P~
exp \{i\int^x_{x_0} B_{\mu}dz_{\mu}\}
 \ee
is the operator of parallel
transport, $\Omega$ is the matrix of global rotations in the color space and
$\bar A_{\mu}(x-x_0)$ in the equation (12)
 \be
 \bar A_{\mu}=\tau^a\bar
\eta^a_{\mu\nu}\frac{x_{\nu}}{x^2} f(x^2)
\ee
 is the instanton field
in some gauge.
The $f$ is so-called "instanton's profile function". Classical function $f$
has the standard form in the singular gauge,
\be
f_{cl}=\frac{1}{1+(x^2/\rho^2)}
\ee
Prior to averaging over the fields $B_{\mu}$, we write down the paramagnetic
term in $S_{eff}$ in the following form\footnote{In this
section the notation $\bar A_{\mu}=A_{\mu}$ for brevity is  used.}
$$
S_{para}{[A,B]}=-\frac{1}{2g^4}\int
d^4xd^4y \int d\Omega tr \Omega F_{\mu\nu}(x) \Omega^+ G_{\mu\nu}(x,x_0)
$$
\be
\times  tr\Omega
F_{\sigma\lambda}(y)\Omega^+ G_{\sigma\lambda} (y, x_0),
\ee
where
\be
G_{\mu\nu}(x,x_0)\equiv \Phi(x_0,x) G_{\mu\nu}(x) \Phi(x,x_0)
\ee
and $d\Omega$ is the Haar measure on the $SU(N_c)$.
In the tensor notation, the integrand in (16) is written as
\be
\Omega_{\alpha_1\beta_1}(F_{\mu\nu})_{\beta_1\alpha_2}
\Omega^+_{\alpha_2\beta_2} (G_{\mu\nu})_{\beta_2\alpha_1}
\Omega_{\alpha_3\beta_3}(F_{\sigma\lambda})_{\beta_3\alpha_4}
\Omega^+_{\alpha_4\beta_4}
(G_{\sigma\lambda})_{\beta_4\alpha_3}
\ee
Taking into account the relation
$$
\int d\Omega\Omega_{\alpha_1\beta_1}\Omega^+_{\alpha_2\beta_2}
\Omega_{\alpha_3\beta_3}\Omega^+_{\alpha_4\beta_4}
$$
 \be
=\frac{1}{N^2_c-1}
\{
\delta_{\alpha_1\beta_2}
\delta_{\beta_1\alpha_2}
\delta_{\alpha_3\beta_4}
\delta_{\beta_3\alpha_4}+
\delta_{\alpha_1\beta_4}
\delta_{\beta_1\alpha_4}
\delta_{\alpha_2\beta_3}
\delta_{\beta_2\alpha_3}\}+ O(\frac{1}{N_c^3})
\ee
and integrating (16) over the Haar measure, one obtains
\be
S_{para}{[A,B]}=-\frac{1}{2(N_c^2-1)g^4}
\int
d^4xd^4ytr F_{\mu\nu}(x)F_{\sigma\lambda}(y)tr G_{\mu\nu}(x,x_0)
G_{\sigma\lambda}(y,x_0).
\ee
Let us perform averaging over the nonperturbative field $B_{\mu}$
and remind that the nonlocal gauge--invariant correlation function
$<tr G_{\mu\nu}(x,x_0)G_{\sigma\lambda}(y,x_0)>$ can be represented
as [1]
\be
 <tr G_{\mu\nu}(x,x_0)G_{\sigma\lambda}(y,x_0)>=
\frac{<G^2>}{12}\{(\delta_{\mu\sigma}
\delta_{\nu\lambda}-\delta_{\mu\lambda}\delta_{\nu\sigma})D(x-y)+O(D_1)\}.
\ee
In this way, the paramagnetic term in $S_{eff}$ is reduced to the
form
\be
S_{para}{[A]}=<S_{para}[A,B]>_B=-\frac{<G^2>}{48(N_c^2-1)g^4}
\int
d^4xd^4y F_{\mu\nu}^a(x)D(x-y)F_{\mu\nu}^a(y).
\ee
Let us consider the diamagnetic part
\be
S_{dia}[A,B]=-\frac{1}{g^2}\int d^4xtr
\{[A_{\mu},B_{\nu}]^2+[A_{\mu},B_{\nu}][B_{\mu},A_{\nu}] \}
.
\ee
Again we  integrate over the Haar
measure
first. Thus, the  $S_{dia}$ is given by the  following expression:
\be
S_{dia}[A,B]
=\frac{3}{2g^2}\frac{N_c}{N_c^2-1}
\int d^4xtr A^2_{\mu} (x) tr B^2_{\mu}(x).
\ee
       To obtain gauge--invariant expressions for vacuum correlator,
       we use the Fock--Schwinger gauge $x_{\mu} B_{\mu} =0$ . By
       virtue of the fact  that the  t'Hooft symbols
       $\bar\eta^a_{\mu\nu}$ are antisymmetric, the instanton field
       satisfies this gauge condition automatically. In this gauge,
       the nonperturbative field $B_{\mu} $ is related to the field
       strength $G_{\mu\nu}$
by the well--known equation
\be
B_{\mu}(x)=x_{\nu}\int^1_0\alpha d\alpha G_{\mu\nu}(\alpha x).
\ee
Using the relation (21) we arrive at
\be
S_{dia}[A] = <S_{dia} [A,B]>_B=\frac{1}{2g^2} \int d^4x\Sigma (x)
(A^a_{\mu}(x))^2,
\ee
where
\be
\Sigma(x) =\frac{3}{16} \frac{N_c}{N_c^2-1}<G^2>x^2\int^1_0 \alpha
d\alpha \int^1_0\beta d\beta D(|\alpha-\beta|x).
 \ee
  Bringing all the
results together, we find that $S_{eff}$ for instanton--like object
in the nonperturbative vacuum is given by
$$
 S_{eff}[A] = S[A]+S_{para}[A]+S_{dia}[A]
 $$
\be
=\frac{1}{4g^2}\int d^4xd^4yF^a_{\mu\nu}(x) \varepsilon (x-y)
F^a_{\mu\nu}(y) + \frac{1}{2g^2} \int d^4x\Sigma (x)
 (A^a_{\mu}(x))^2,
 \ee
where
 \be
 \varepsilon (x-y) =\delta^4(x-y)-\Pi(x-y),
 ~~\Pi(x-y) =\frac{<G^2>}{12(N_c^2-1)g^2}D(x-y)
 \ee
 Equation of motion for $A_{\mu}$ is
\be
\nabla_{\mu}^{ab}(A(x))\int d^4 y \varepsilon (x-y)
F^b_{\mu\nu}(y)-\Sigma(x) A^a_{\nu} (x) =0
 \ee
and
$$
\nabla_\mu^{ab}=\delta^{ab}\partial_{\mu}-if^{abc}A^c_\mu
$$
is the covariant derivative on the gauge group $SU(N_c)$.

\section{Scale transformations}

In the previous  section  we have derived $S_{eff}$ for  weak external field
(instanton--like object of large size) in
the nonperturbative gluon vacuum. It is known that
scale invariance in gluodynamics is broken due to the anomaly of the trace of
energy--momentum tensor  $<\theta_{\mu\mu}>\sim -<G^2>$. Topology
allows the existence of topologically  nontrivial solutions. However,
their stability in size is not evident from scale arguments. For
example, in theories which include Higgs fields (Weinberg--Salam or
Georgi--Glashow models) instanton solutions are absent at the
classical level. Instanton tends to contract in size thus lowering
the classical action and finally becoming a point singularity with
$\rho=0$.

If however one considers the effective action $S_{eff}$ with quantum
corrections over the instanton background field, one   can perform a scale
transformation
$x_\mu\to x_\mu/\lambda,~~A_\mu(x)\to\lambda^{-1}A_\mu(x/\lambda)$
 and then consider the derivative of $S_{eff}$ with respect
of the logarithm of the transformation parameter. This yields
 \be
\frac{\partial S_{eff}}{\partial ln\lambda}\Biggl |_{\lambda=1}=
\int d^4x\theta_{\mu\mu}(x)\Biggr. \;\; ,
\ee
where $\theta_{\mu\mu}(x)$ is the trace of energy--momentum tensor for a
given field configuration. The trace $\theta_{\mu\mu}(x)$ contains a
positively defined contribution which is due to the mass term of the
classical action. This mass term breaks scale invariance and leads to the
collapse of the classical instanton. In the quantum theory
$\theta_{\mu\mu}$  acquires a negative contribution due to quantum anomaly.
It is the presence of this anomaly which eventually stabilizes the instanton.
One may argue that the size of the instanton is determined from the condition
of  the total pressure to be zero
$$
\int d^4x\theta_{\mu\mu}(x) =0.
$$

Let us consider the scale transformation of the effective action (28) and
prove the existence of the stable in size instanton in nonperturbative scale
noninvariant gluodynamics.

Numerical lattice calculations revealed that, to a high accuracy, the function
$D$ can be approximated as [3,4]
\be
D(x-y)=e^{-\mu|x-y|},~~ \mu\equiv 1/T_g\approx 1 GeV.
\ee
Let write (22) in the Fourier form
\be
S_{para}[A]=-\frac{1}{4g^2}\int\frac{d^4q}{(2\pi)^4}F^a_{\mu\nu}(q)
F^a_{\mu\nu}(-q)\Pi(q^2).
\ee
Taking into account (29) one obtains
\be
\Pi(q^2)=\frac{<G^2>}{12(N^2_c-1)g^2}
\int d^4xe^{iqx}D(x)
=\frac{<G^2>}{(N^2_c-1)g^2}\frac{\pi^2\mu}{(\mu^2+q^2)^{5/2}}.
\ee
Perform now a scale transformation $q_\mu\to \lambda q_{\mu},~~
F^a_{\mu\nu}(q)\to \lambda^2 F^a_{\mu\nu} (\lambda q)$. Then
\be
\frac{\partial S_{para}}{\partial ln \lambda}
\Biggl |_{\lambda=1}=
\frac{1}{2g^2}\int \frac{d^4q}{(2\pi)^4} F^a_{\mu\nu}(q) F^a_{\mu\nu}(-q)
\frac{\partial \Pi (q^2)}{\partial ln q^2} < 0 \Biggr. .
\ee
Consider $S_{dia}$ in the same way. Performing a transformation
 $x_\mu\to x_{\mu/}\lambda$,
$~~A_\mu(x)\to {\lambda}^{-1} A_\mu({x}/{\lambda})$
one arrives at
\be
\frac{\partial S_{dia}}{\partial ln \lambda}\Biggl |_{\lambda=1}=
\frac{1}{g^2}\int d^4 x (A^a_\mu(x))^2(\Sigma(x)+\frac{\partial
\Sigma(x)}{\partial ln x^2}) \Biggr. .
\ee
 Making use of the relations (27) and (32) and of the conclusions on\\
$D(|\alpha-\beta|x)$ from [9] one can deduce that $\Sigma
(x)+\partial \Sigma (x)/\partial ln x^2>0$.  Therefore
 \be
 \frac{\partial
S_{dia}}{\partial ln \lambda}\Biggl |_{\lambda=1} > 0 \Biggr. .
  \ee
 It means
that diamagnetic interaction of the instanton with nonperturbative fields
leads to the collapse of the instanton. On the other hand the paramagnetic
interaction  effectively results in instanton "swelling" in its scale. Hence
a stable in size instanton can exist in the nonperturbative gluonic vacuum
described by nonlocal gauge--invariant correlator $<tr
G_{\mu\nu}(x,x_0)G_{\sigma\lambda}(y,x_0)>.$

\section{Screening}

The instanton profile function in
the  nonperturbative vacuum is determined by Eq.
(30) for the ansatz (14). At small distances from the center of the
instanton $x\to 0, (q\to \infty)$ equations (27), (29) and (34) lead to
\footnote{In this section, we use the notation $x=\sqrt{x^2_\mu}$.}
 $$
\Sigma (x\to 0)\to \frac{1}{16} \frac{N_c}{N_c^2-1}<G^2>x^2,
$$
\be
\varepsilon (q\to \infty) \to 1
\ee
So at small distances one retrieves the classical equation for the
profile function
\be
\frac{d^2f}{dx^2}+2\frac{df}{dx}
-\frac{8}{x}(2f^3-3f^2+f)=0,~~ f(0)=1,f(\infty)=0
\ee
 and, correspondingly,
$f(x\to 0)\to f_{cl}(x)$ (15).\\
At large distances $x\to \infty,(q\to 0)$
$$
\Sigma (x\to \infty)\simeq \frac{1}{8} \frac{N_c}{N_c^2-1}<G^2>
T_gx (1+O(\frac{T_g}{x}))
$$
 \be
\varepsilon (q\to 0) \simeq
1-\frac{\pi^2}{(N_c^2-1)g^2}<G^2> T_g^4
 \ee
 Thus, we have the
following asymptotic equation for the profile function \be
\frac{d^2f}{dx^2}-\gamma xf=0,~~ f(\infty)=0,
\ee
where
\be
\gamma=\frac{1}{8}\frac{N_c}{N_c^2-1}<G^2>T_g(1
-\frac{\pi^2}{(N_c^2-1)g^2}<G^2> T_g^4 )^{-1}
\ee
 Hence the function $f$ is asymptotically damped at large distances like
the Airy  function
 \be
f(x)\sim Ai(\gamma^{1/3}x)\to \frac{1}{2}
\sqrt{\frac{1}{\pi}}\gamma^{-1/12}x^{-1/4}
exp(-\frac{2}{3}\gamma^{1/2}x^{3/2}),~~ x\to \infty
 \ee
One notice that from (43) we have screening of the instanton profile
function with effective mass $m_{sc}=(2/3)^{2/3}\gamma^{1/3}\approx
\gamma^{1/3}$. Choosing the standard values of the parameters in QCD  $N_c=3,
T_g=1 GeV^{-1}, <G^2>=0,5 GeV^4, \alpha_s=g^2/4\pi =0,3$ one obtains
the  screening mass
$m_{sc}(QCD)\simeq 250 MeV$.
On the other hand, as it is well--known, in gluodynamics the value of the
vacuum condensate is $<G^2>_{GD}\simeq  4 <G^2>$ [13]. Therefore, one obtains
the screening mass $m_{sc}(GD)\simeq 520 MeV$.

\section{Conclusion}

Thus, making use of Vacuum Correlators Method we have obtained the effective 
action for the topologically nontrivial gluon field fluctuation 
(instanton--like object) interacting with stochastic vacuum fields.
Effective action appears to be essentially nonlocal. This nonlocal
interaction (paramagnetic effect) leads eventually to the existence of the 
instanton which is scale noncollapsing.

It should be stressed that we have used exponentially decreasing ansatz (32)
for the D--function. However, it can be easily seen that our result is
valid for any function D that decrease sufficiently fast. The only
difference is in the numerical value of $\gamma$. For example, for 
Gaussian
ansatz one has $D(z)=\exp(-\mu^2z^2)$, $\gamma\to C\cdot\gamma$ and
$C\approx0.9$.

One concludes that the effect of screening is able to change drastically the
interaction in the instanton liquid (dipole--dipole attraction at the large
distances), where the average distances between pseudoparticles ($\sim
(200
MeV)^{-1}$) are larger than the inverse screening mass
($\sim (500 MeV)^{-1}$).  The numerical calculations for the
instanton screening effects in the nonperturbative gluodynamics vacuum and
comparison with the lattice calculations are in progress.

\noindent
{\large \bf Acknowledgments}

\noindent
The author is grateful to Yu.~S.~Kalashnikova, B.~O.~Kerbikov and
Yu.~A.~Simonov for useful discussions. The work was supported in part by
RFFI grant ü 96-02-19184a and INTAS grant ü 94-2851.


\begin{thebibliography}{99}
\bibitem{1}
H.~G.~Dosch, Phys. Lett. {\bf B190}, 177 (1987);
H.~G.~Dosch, and Yu.~A.~Simonov, ibid {\bf B205}, 339 (1988);
 Yu.~A.~Simonov, Nucl. Phys. {\bf B324}, 67 (1989).
\bibitem{2}
 Yu.~A.~Simonov, Yad. Fiz. (Sov. Phys.) {\bf 54}, 192 (1991);
 Yu.~A.~Simonov, Usp. Fiz. Nauk {\bf 166}, 67 (1996), and references therein.
\bibitem{3}
M.~Campostrini, A.~Di~Giacomo and G.~Mussardo, Z. Phys. {\bf C25}, 173
(1984);
 A.~Di~Giacomo and H.~Panagopoulos, Phys. Lett. {\bf B285}, 133 (1992);
L.~Del~Debbio,
 A.~Di~Giacomo and Yu.~A.~Simonov, Phys. Lett.  {\bf B332}, 111 (1994).
\bibitem{4}
 A.~Di~Giacomo, E.~Meggiolaro and
 H.~Panagopoulos, Nucl. Phys. Proc. Suppl. {\bf 54 A}, 343 (1997);
 A.~Di~Giacomo, E.~Meggiolaro and
 H.~Panagopoulos, Nucl. Phys. {\bf B483}, 371 (1997).
 \bibitem{5}
 E.~V.~Shuryak, Nucl. Phys.  {\bf B203}, 93 (1981)
 \bibitem{6}
 D.~I.~Diakonov and V.~Yu.~Petrov, Nucl. Phys. {\bf B245}, 259 (1984).
 \bibitem{7}
T.~Schaefer and E.~V.~Shuryak, Rev. Mod. Phys. {\bf 70}, 323 (1998), and
references therein.

 \bibitem{8}
 N.~O.~Agasian and Yu.~A.~Simonov, preprint ITEP-78-94, Mod.Phys. Lett.
 {\bf A10}, 1755 (1995).
 \bibitem{9}
 N.~O.~Agasian, Phys. Atom. Nucl. {\bf 59}, 297 (1996).
 \bibitem{10}
 A.~B.~Migdal, N.~O.~Agasian and S.~B.~Khokhlaclev, JETF Lett. {\bf 41}, 497
(1985);
 N.~O.~Agasian and S.~B.~Khokhlaclev, Sov. J. Nucl. Phys. {\bf 55}, 628 (1992);
N.~O.~Agasian and S.~B.~Khokhlaclev, ibid. {\bf 55}, 633 (1992).
 \bibitem{11}
M.~A.~Shifman, A.~I.~Vainshtein and V.~I.~Zakharov, Nucl. Phys. {\bf B147}, 385
(1979).
 \bibitem{12}
A.~M.~Polyakov, Gauge Fields and Strings (Harwood Academic, Char, Switzerland,
1987).
 \bibitem{13}
 A.~I.~Vainshtein, V.~I.~Zakharov, V.~A.~Novikov and M.~A.~Shifman, Physics of
Elementary particles and atomic nuclei,   {\bf 13}, 542 (1982).
\end{thebibliography}
\end{document}